\documentclass[prb,aps,showpacs,twocolumn,unsortedaddress]{revtex4-1}
\usepackage{graphicx,bm}
\usepackage{amssymb}
\usepackage{amsmath}
\usepackage{epsfig}
\usepackage{epsf}
\usepackage[usenames]{color}
\usepackage{hyperref}

\begin{document}

\title{Current-induced torques in textured Rashba ferromagnets}

\author{E. van der Bijl}
\email[Electronic address: ]{e.vanderbijl@uu.nl}
\author{R.A. Duine}
\affiliation{Institute for Theoretical Physics, Utrecht
University, Leuvenlaan 4, 3584 CE Utrecht, The Netherlands}

\date{\today}
\begin{abstract}
In systems with small spin-orbit coupling, current-induced torques on the magnetization require inhomogeneous magnetization textures. For large spin-orbit coupling, such torques exist even without gradients in the magnetization direction. Here, we consider current-induced torques in ferromagnetic metals with both Rashba spin-orbit coupling and inhomogeneous magnetization. We first phenomenologically construct all torques that are allowed by the symmetries of the system, to first order in magnetization-direction gradients and electric field. Second, we use a Boltzmann approach to calculate the spin torques that arise to second order in the spin-orbit coupling. We apply our results to current-driven domain walls and find that the domain-wall mobility is strongly affected by torques that result from the interplay between spin-orbit coupling and inhomogeneity of the magnetization texture.
\end{abstract}
\maketitle

\def\bx{{\bm x}}
\def\bX{{\bm X}}
\def\bk{{\bm k}}
\def\bK{{\bm K}}
\def\bq{{\bm q}}
\def\bv{{\bm v}}
\def\bxi{{\bm \xi}}
\def\bma{{\bm m}}
\def\br{{\bm r}}
\def\bp{{\bm p}}
\def\bpi{{\bm \pi}}
\def\bM{{\bm M}}
\def\bs{{\bm s}}
\def\bS{{\bm S}}
\def\bB{{\bm B}}
\def\bE{{\bm E}}
\def\bA{{\bm A}}
\def\bj{{\bm j}}
\def\bF{{\bm F}}
\def\uz{{\bm e}_z}
\def\fe{{\mathcal F}}

\def\id{{\rm d}}
\def\bOm{{\bm \Omega}}
\def\bH{{\bm H}}

\def\br{{\bm r}}
\def\bv{{\bm v}}

\def\half{\frac{1}{2}}
\def\args{(\bm, t)}

\def\rdw{r_{\rm{dw}}}
\def\phidw{\varphi_{\rm{dw}}}
\def\ldw{\lambda_{\rm{dw}}}
\def\thetadw{\theta_{\rm{dw}}}

\section{Introduction}
Current-induced torques on the magnetization in conducting ferromagnets are one of the main topics of research in spintronics. In addition to being fundamentally interesting, these torques are also key to developments in memory technology.\cite{Parkin2008} Current-induced torques can be used to move domain walls through a ferromagnetic wire. When a domain wall is present the direction of the magnetization depends on the position in the wire. This spatial dependence of the magnetization gives rise to a mismatch between the electron spin polarization and local magnetization resulting in the adiabatic reactive \cite{Slonczewski1996,Berger1996} and dissipative (also known as non-adiabatic) spin transfer torques (STTs).\cite{ZhangLi2004,BarnesMaekawa2005,TserkovnyakSkadsemBrataasBauer2006,KohnoTataraShibata2006,PiechonThiaville2007,DuineNunezSinovaMacDonald2007,*Duine2009} The occurence of these two spin torques is well estabished but their relative magnitude, parametrized by the dimensionless parameter $\beta$ which describes the relative strength of the dissipative torque with respect to the reactive one, is hard to measure\cite{ThomasParkin2006,*HeyneMattheis2008,*BolteStoll2008,*HeyneKronast2010} and calculate.\cite{LucassenWongDuineTserkovnyak2011}

That there exist other current-induced torques related to spin-orbit (SO) coupling of the cariers has been proposed recently.\cite{ManchonZhang2008,*ManchonZhang2009,ObataTatara2008,GarateMacDonald2009,Hals2009} In these works systems with SO coupling and homogeneous magnetization are considered. Recent experiments can be interpreted using these current-induced torques originating from the SO coupling of the carriers\cite{Miron2011,*MironNmat2011,Liu2011,Kim2012,PesinMacDonald2012,Ryu2012} that, unlike the adiabtic STT mentioned above, do not require magnetization gradients. (Note, however, that these observations can also be described via the Spin-Hall effect in Pt as argued in Ref. [\onlinecite{Liu2011}].) For Rashba SO couping two current-induced torques have been found. In the experimental works a domain wall is present. This implies that the description in terms of a homogeneous magnetization is incomplete and a more systematic description including both SO coupling and an inhomogeneous magnetization is called for.

It is the purpose of this work to give such an inclusive description that incorporates both SO coupling and inhomogenous magnetization textures. For definiteness, we focus on the Rashba SO coupling. In Sec. \ref{sec:SymCon} we consider all current-induced torques which are allowed by the symmetries of the system. As the number of allowed torques is considerable, and because the symmetry considerations do not yield their relative magnitudes, we investigate these within a semi-classical Boltzmann description. In Sec. \ref{sec:DWM} the results for the torques are used to calculate their effect on domain-wall dynamics. We find that the current-induced domain-wall velocity depends strongly on wall geometry. Furthermore, the domain-wall mobility depends strongly on the inclusion of torques that result from the interplay of SO coupling and gradients in the magnetization.
\section{Symmetry Considerations}\label{sec:SymCon}
In this section we use symmetry considerations to obtain all allowed current-induced torques. To illustrate our method we begin with the adiabatic spin torques in the absence of SO coupling. Subsequently we investigate the situation with SO coupling. We use the $s-d$ model since this is a convenient model to get the qualitative description of current-induced torques. In this model the magnetization resides on the $d$-orbitals and transport is due to the mobile $s$-electrons. We investigate the system well below the Curie temperature, which means the magnetization is represented using a unit-vector field since fluctuations in its magnitude are negligible.
\subsection{Absence of Spin-Orbit Coupling}
Within the $s-d$ model the Hamiltonian is given by
\begin{equation}\label{eq:Hsd}
\mathcal{H}_{sd} =\mathcal{H}_{0}(\bx,\bp) -\frac{\Delta}{2}\bma\cdot\bs,
\end{equation}
where $\mathcal{H}_{0}$ is the Hamiltonian that descibes the motion of the itinerant electrons and depends on electron momentum $\bp$ and postion $\bx$. We have an exchange coupling between the magnetic texture $\bma(\bx,t)$ and the electron spin $\bs(t)$ specified by the exchange splitting $\Delta$. The total Hamiltonian $\mathcal{H}_{sd}$ is invariant under two \textit{independent} rotations of the spin and physical space, parameterized by the rotation matrices $\mathcal{R}^{ij}_{\rm{S}}$ and $\mathcal{R}^{ij}$ respectively. (We neglect the coupling between the magnetization and the orbit of the electrons that occurs via the Lorentz force. We neglect this effect for the moment because the magnetic field induced by the magnetization is very small.) Moreover, in this description we neglect the ionic lattice. We explicitly have for the rotations
\begin{eqnarray}
\tilde{s}^i &=& \mathcal{R}^{ij}_{\rm{S}} s^j, \qquad \tilde{m}^i = \mathcal{R}^{ij}_{\rm{S}} m^j;\\
\tilde{x}^i &=& \mathcal{R}^{ij} x^j, \qquad \tilde{p}^i= \mathcal{R}^{ij} p^j.
\end{eqnarray}
Note that we use the summation convention of summing over repeated indices. The invariance of the Hamiltonian implies $\mathcal{H}_{sd}(\tilde{\bp},\tilde{\bx},\tilde{\bs},\tilde{\bma})=\mathcal{H}_{sd}(\bp,\bx,\bs,\bma)$. This means that these symmetries should be respected at the level of the equations of motion. We are interested in the (linear-response-) current-induced torques, hence our expressions for the torques should be linearly dependent on the applied electric field $\bE$. The possible torques that are first order in the electric field $\bE$, which transforms under the action of $\bm{R}$, should involve an inner-product with another vector that transforms under the same rotation and in this way creates an invariant scalar. The only other vector that transforms in this way for this system is the gradient $\nabla$ that acts on the magnetization. These constraints lead to the two possible current-induced torques 
\begin{equation}\label{eq:adiabatictorques}
\left.\frac{\partial \bma}{\partial t}\right|_{\rm{ST}} \propto (\bE \cdot \nabla)\bma +\beta \bma\times(\bE \cdot \nabla)\bma.
\end{equation}
For a treatment of spin transfer torques that incorporates the symmetries of the lattice see Ref. [\onlinecite{HalsBrataas2012}]. These terms are frequently written in terms of the current but we choose to put in the electric field here as the external perturbation, to be consistent with the rest of this paper. Note the parameter $\beta$ which is defined as the ratio of the dissipative and reactive spin transfer torques. 

The two torques in Eq. (\ref{eq:adiabatictorques}) are mutually perpendicular. Moreover they transform differently under time reversal, since they differ by a factor $\bma$ which is odd under time-reversal. This difference in behaviour under time-reversal symmetry implies the torques form a pair where one is reactive and the other is dissipative.
\subsection{Spin-Orbit Coupling}\ref{sec:SymConSOC}
In the presence of SO coupling the Hamiltonian for the spin of the $s$-electron couples the spin and the momentum of the electron. We represent SO coupling for spin-$\frac{1}{2}$ carriers via the Hamiltonian
\begin{equation}
\mathcal{H}_{\rm{SO}}=-\bOm(\bx,\bp)\cdot\bs,
\end{equation}
where $\bOm$ contains both the exchange interaction of Eq. (\ref{eq:Hsd}) and SO coupling, and can be seen as a position and momentum dependent effective exchange splitting.

For definiteness, and motivated by experiments,\cite{Miron2011,*MironNmat2011,Liu2011} we study the simplest form of SO coupling described by the Rashba Hamiltonian\cite{Rashba1960} $\mathcal{H}_{\rm{R}}=-\lambda_R (\bp\times\uz)\cdot\bs$. The Rashba coupling together with the exchange interaction results in
\begin{equation}\label{eq:bOm}
\bOm(\bx,\bp)=\frac{\Delta}{2}\bma(\bx)+\lambda_{\rm{R}}\bp\times\uz.
\end{equation}
Rashba SO coupling occurs in two-dimensional electon systems with inversion asymmetry along the direction perpendicular to the two-dimensional electron gas (which we choose as our z-axis). The SO coupling breaks the invariance of the Hamiltonian under separate rotations of the spin and orbital parts of the motion. Total angular momentum is still conserved due to the invariance of the Hamiltonian under combined rotations of spin and physical space, parameterized by $\mathcal{R}^{ij}_{\rm{S}}= \mathcal{R}^{ij}$. 

The linear-response matrix $L_{\rm{cit}}(\bma,\uz,\nabla\bma)$ that describes the current-induced torques is defined by
\begin{equation}\label{eq:Lcit}
 \dot{m}^i = L_{\rm{cit}}^{ij}(\bma,\uz,\nabla\bma)  E^j,
\end{equation}
 where $\bE$ is the electric field in the plane and $\uz$ is a unit vector in the z-direction. The linear-response matrix depends on this direction since inversion symmetry is broken along this direction. The Hamiltonian is invariant under parity transformations which implies that the linear response matrix should obey
$-L_{\rm{cit}}(\bma,-\uz,-\nabla\bma)=L_{\rm{cit}}(\bma,\uz,\nabla\bma)$. This shows that there can be torques on the magnetization without a gradient in the magnetization. These torques $\tau_{\rm{ST}} i=L_{ij}(\bma,\uz) E^j$ have been found before\cite{ManchonZhang2009} and are given by
\begin{eqnarray}
\bm{\tau}^{(1)}_{\rm{ST}} &\propto& \bma\times(\bE\times\uz);\label{eq:tst1}\\
\bm{\tau}^{(1\perp)}_{\rm{ST}}&\propto& \bma\times( \bma\times(\bE\times\uz) ).\label{eq:tst1p}
\end{eqnarray}
The spin torques are perpendicular to $\bma$ because it is a unit vector field. Since the magnetization is embedded in three-dimensional space there is a two-dimensional plane perpendicular to it. This means that any spin torque $\bm{\tau}^{(i)}_{\rm{ST}}$ allowed by the symmetry of the system immediately defines another torque via $\bm{\tau}^{(i\perp)}_{\rm{ST}}=\bma\times\bm{\tau}^{(i)}_{\rm{ST}}$. These pairs differ a factor $\bma$ which changes its sign under time-reversal, hence the two torques form a reactive-dissipative pair, like the STTs in Eq. (\ref{eq:adiabatictorques}). In the following we will show only one of the pair. All terms to first order in the gradient of the magnetization that do not involve $\uz$, are given by
\begin{eqnarray}
\bm{\tau}^{(2)}_{\rm{ST}} &\propto& (\bE\cdot\nabla)\bma;\\
\bm{\tau}^{(3)}_{\rm{ST}} &\propto& ((\bma\times\bE)\cdot\nabla)\bma;\\
\bm{\tau}^{(4)}_{\rm{ST}} &\propto& (\bma\cdot\bE)(\bma\cdot\nabla)\bma;\\
\bm{\tau}^{(5)}_{\rm{ST}} &\propto& E^a(\bma\times\nabla)\bma^a;\\
\bm{\tau}^{(6)}_{\rm{ST}} &\propto& (\bma\times\bE)^a(\bma\times\nabla)\bma^a;\\
\bm{\tau}^{(7)}_{\rm{ST}} &\propto& \bma\times\bE(\nabla\cdot \bma);\label{eq:CIt7}\\
\bm{\tau}^{(8)}_{\rm{ST}} &\propto& (\bma\times\bE)\bma\cdot(\nabla \times \bma),
\end{eqnarray}
In the first line the familiar STT\cite{Berger1996,Slonczewski1996} describing the current-induced torque in systems with inhomogeneous magnetization is obtained. Together with the dissipative STT\cite{ZhangLi2004,BarnesMaekawa2005,TserkovnyakSkadsemBrataasBauer2006,KohnoTataraShibata2006,PiechonThiaville2007,DuineNunezSinovaMacDonald2007,*Duine2009} that is associated with it ($\tau_{\rm{ST}}^{2\perp}$) those torques describe the weak SO coupling situation. In the second line we find a STT due to a Hall current. The other torques do not have a straightforward physical interpretation.

Up to this point we have explicitly given the torques to first order in either $\uz$ or $\nabla$. There are more torques that involve an even number of $\uz$'s and are first order in $\nabla$. We will not list them because the list will be to long to be illuminating. We proceed by actually calculating the torques in the next section. The reason we do this is twofold. First, having demonstrated the existence of many spin torques due to the combined effects of SO coupling and magnetization gradients, we now explicitly calculate which torques occur within a semi-classical approach to the Rashba model. The second reason is to give an estimate of the relative magnitude of the various current-induced torques which cannot be found using symmetry arguments.
\section{Semi-classical framework}\label{sec:SemiClassDes}
In order to investigate microscopically which current-induced torques appear for the textured Rashba ferromagnet we use a semi-classical approach. This approach has proved its merit in the description of the anomalous Hall effect.\cite{Sinitsyn2008,*Jungwirth2002,*vanderBijl2011}
We describe the system by the Hamiltonian
\begin{equation}
\mathcal{H} = \frac{\bp^2}{2 m_e}-\bOm(\bx,\bp)\cdot\bs+E_{\rm{MM}}\left[\bma\right],
\end{equation}
where $\bOm(\bx,\bp)$ is the effective Zeeman field, given in Eq. (\ref{eq:bOm}), that incorporates the Rashba SO coupling and the exchange coupling, and $E_{\rm{MM}}\left[\bma\right]$ is the micromagnetic energy functional for the magnetization. Furthermore, $m_e$ is the effective mass of the electron. The equation of motion for the spin degree of freedom is written as
\[
\frac{d \bs}{d t} = \frac{1}{\hbar}\bs\times\bOm-\frac{\alpha}{\hbar}\bs\times(\bs\times\bOm),
\]
where we introduced a damping term proportional to $\alpha$ that describes relaxation of the spin into the direction of the effective Zeeman field. The spin dynamics is much faster than the motion of the electrons such that we can solve the above equation of motion up to first order in time derivatives of $\bOm$. We obtain the following solutions
\begin{equation}\label{eq:1storderspindyn}
\bs_s = s \hat{\bOm} +s\frac{\hbar}{\sqrt{\bOm\cdot\bOm}} \frac{d\hat{\bOm}}{dt}\times\hat{\bOm} - \frac{\hbar\alpha}{\sqrt{\bOm\cdot\bOm}} \frac{d\hat{\bOm}}{dt},
\end{equation}
where $s=\pm1$ describe the majority($s=+$)/minority($s=-$) electrons, and $\hat{\bOm}=\bOm/|\bOm|$. The first term describes the adiabatic following of the effective magnetization texture by the electron-spins. The other terms describe the slight mismatch of the spins with the effective magnetization. We find the dynamics of the itinerant electrons by inserting the first order solutions of the spin degree of freedom, given in Eq. (\ref{eq:1storderspindyn}), into the Hamilton equations of motion for the electrons. We obtain
\begin{eqnarray*}
\dot{x}^i_s &=&\frac{\partial \epsilon_{s}}{\partial p^i} -s\hbar \left( \frac{\partial \hat{\bOm}}{\partial p^i}\times\frac{d \hat{\bOm}}{dt} \right)\cdot\hat{\bOm}+\alpha\hbar\frac{\partial\hat{\bOm}}{\partial p^i}\cdot\frac{d\hat{\bOm}}{dt};\\
\dot{p}^i_s &=&-\frac{\partial \epsilon_{s}}{\partial x^i} +s\hbar \left( \frac{\partial \hat{\bOm}}{\partial x^i}\times\frac{d \hat{\bOm}}{dt} \right)\cdot\hat{\bOm}-\alpha\hbar\frac{\partial\hat{\bOm}}{\partial x^i}\cdot\frac{d\hat{\bOm}}{dt}-|e|\bE,
\end{eqnarray*}
where $\epsilon_{s}=\bp^2/2m_e-s|\bOm|$ is the dispersion for the majority($s=+$)/minority($s=-$) electrons. Note that we added an electric field to induce a transport current. The total time derivatives on $\hat{\bOm}$ should be understood as
\[
\frac{d\hat{\bOm}}{d t} = \dot{x}^i_s \frac{\partial \hat{\bOm}}{\partial x^i}+\dot{p}^i_s \frac{\partial \hat{\bOm}}{\partial p^i}.
\]
Now that we have this semi-classical description of the single particle dynamics we calculate the spin-torques using the Boltzmann equation for the distribution function $f_s(\bx,\bp,t)$ of the particles, which, in the relaxation-time ($\tau_r$) approximation, is given by,
\begin{equation}\label{eq:Boltzmanneq}
\frac{d}{dt}f_s(\bx,\bp,t)= -\frac{f_s(\bx,\bp,t)-f^{\rm{FD}}(\epsilon_s)}{\tau_r},
\end{equation}
where $f^{\rm{FD}}(\epsilon)=(1+e^{\beta \epsilon})^{-1}$ is the Fermi-Dirac distribution function. The relaxation-time approximation is the simplest description of the Boltzmann collision integral. We make the relaxation-time approximation here for convenience. A detailed study of the collision-integral in the presence of strong SO coupling is beyond the scope of this work. We refer to the work by Pesin and MacDonald in Ref. [\onlinecite{PesinMacDonald2012}] for more details on the situation of homogeneous magnetization. The left-hand side in Eq. (\ref{eq:Boltzmanneq}) should be read as
\[
\frac{d f_s}{dt}=\frac{\partial f_s(\bx,\bp,t)}{\partial \bma}\cdot\dot{\bma}+ \frac{\partial f_s(\bx,\bp,t)}{\partial \bp}\cdot\dot{\bp}_s+\frac{\partial f_s(\bx,\bp,t)}{\partial \bx}\cdot\dot{\bx}_s.
\]
The equation of motion for the magnetization is the Landau-Lifshitz-Gilbert (LLG) equation
\begin{equation}\label{eq:LLG}
\frac{\partial \bma}{\partial t} = -\gamma\bma\times \bH_{\rm{eff}}+\alpha_G \bma\times\frac{\partial \bma}{\partial t}+\bm{\tau}_{sd},
\end{equation}
where $\gamma$ is the gyromagnetic ratio and the torques due to the $s-d$ coupling $\bm{\tau}_{sd}=\Delta/(2\hbar) \bma\times\langle\bs\rangle$ contain the spin torques of interest and a renormalization of the parameters in the LLG\cite{ZhangLi2004} equation which we discuss in this section. The current-induced torques are proportional to the electric field and will be given in Sec. \ref{sec:CItorques}. The renormalized LLG equation we obtain is given by
\begin{equation}\label{eq:LLGr}
(1-\eta)\frac{\partial \bma}{\partial t} =-\gamma \bma\times\bH'_{\rm{eff}}+\alpha'_G \bma\times\frac{\partial \bma}{\partial t}+{\bm \tau}_{\rm{ST}},
\end{equation}
where $\bm{\tau}_{\rm{ST}}$ contains all terms of $\bm{\tau}_{\rm{sd}}$ proportional to the electric field and $\bH'_{\rm{eff}}$ is defined as the effective magnetic field acting on the magnetization which acquires an additional term from the coupling to the electrons
\begin{equation}\label{eq:Heffprime}
\bH'_{\rm{eff}}=\frac{\delta E_{\rm{MM}}}{\delta\bma}+\alpha\frac{m_e^2 a^2\lambda_R^2}{2\gamma \pi\hbar^2}(1+\frac{4\epsilon_F^2}{\Delta^2})(\dot{\bma}\cdot\uz)\uz,
\end{equation}
and the renormalized quantities in Eq. (\ref{eq:LLGr}) are given by
\begin{equation*}
\eta = \frac{m_e a^2}{\pi \hbar^2}\left(\frac{\Delta}{2} - 4\frac{m_e \lambda^2\epsilon_F}{\Delta}m_z^2  \right);
\end{equation*}
where $a$ is the lattice constant. The additional term in Eq. (\ref{eq:Heffprime}) is an anisotropic damping term which for the typical parameters (see Table \ref{tbl:parameters}) we use in the calculation of the domain wall dynamics is negligible, moreover these parameters also imply $\eta\ll 1$. Furthermore we obtain that the observed Gilbert damping constant is given by
\begin{equation}
\alpha_G'= \alpha_G-\alpha\frac{a^2 m_e}{\pi \hbar^2}\left(\epsilon_F -\frac{m_e\lambda^2}{2}(1-\frac{4\epsilon_F^2}{\Delta^2}(1+4m_z^2))\right).
\end{equation}
Note that $\alpha_G$ phenomenologically describes the damping of the magnetzation due to interactions other than the $s-d$ coupling, such as relaxation due to magnon-phonon interactions. Before we calculate the spin torques within this semi-classical framework we determine the current as a function of electric field within this simple model. We need this  later on to express the spin torques in terms of the current.
\section{Conductivity}\label{sec:Conductivity}
In this section we give the conductivity for the Rashba system. Note that the conductivity we find here is only correct within this simple $s-d$ description. We need the conductivity in order to interpret the current-induced torques in the next section. The conductivity $\sigma_{ij}$ is defined as $j_c^i=\sigma^{ij}E^i$, where $\bj_c$ is the charge-current density. We calculate the conductivity up to first order in the gradient of the magnetization and up to second order in the SO coupling strength. The expression for the charge-current density is given by
\begin{equation}
\bj_c=-|e|\sum_{s=\pm}\int\frac{d^2\bp}{(2\pi\hbar)^2} f_s(\bx,\bp,t)\dot{\bx}_s.
\end{equation}
Using the relaxation-time approximation described in the previous section we find that the conductivity has three contributions $\sigma=\sigma_0 + \sigma_{\rm{AH}}+\sigma_{\rm{AMR}}$ corresponding to the diagonal, anomalous Hall effect and anisotropic magnetoresistance, respectively. The diagonal conductivity is given by
\begin{equation}\label{eq:sigma0}
\frac{\sigma^{ij}_0}{G_0}=\left(\frac{\epsilon_F\tau_r}{\hbar}+\alpha\frac{4m_e\lambda^2 \epsilon_F}{\Delta^2} \right) \delta^{ij},
\end{equation}
where $G_0=2|e|^2/h$ is the quantum of conductance. The second contribution
\begin{equation}
\frac{\sigma^{ij}_{\rm{AH}}}{G_0}=\Bigg(\frac{2m_e\lambda^2}{\Delta}m_z+\lambda\tau_r( \bma\cdot(\nabla\times\bma)-\alpha\frac{2\epsilon_F}{\Delta}(\nabla\cdot\bma))\Bigg) \epsilon^{ijz},
\end{equation}
which is the anomalous Hall response generalized to inhomogeneous magnetization. The last contribution to the conductivity is 
\begin{equation}\label{eq:sigmaAMR}
\frac{\sigma^{ij}_{\rm{AMR}}}{G_0}=m_e \lambda^2\left(\alpha\frac{4\epsilon_F}{\Delta^2}-\frac{\tau_r}{\hbar} \right)\epsilon^{i a z}\epsilon^{j b z}m^{a}m^{b},
\end{equation}
which depends on the relative orientation of the electric field and the magnetization and hence corresponds to anisotropic magnetoresistance. We also define the current polarization via
\[
\mathcal{P}\bj_c \equiv -|e|\sum_{s=\pm}\int\frac{d^2\bp}{(2\pi\hbar)^2} f(\epsilon_s)s\dot{\bx}_s,
\]
for later reference.
\section{Current-Induced Torques}\label{sec:CItorques}
In this section we give the current-induced torques for the Rashba model, introduced in Sec. \ref{sec:SemiClassDes}. The current-induced torques can be calculated from the current-induced spin density. They are given by
\begin{eqnarray}
\bm{\tau}_{\rm{ST}} &=& \frac{\Delta}{2\hbar}\bma\times\langle\bs\rangle\nonumber\\
&=&\frac{\Delta a^2}{2\hbar}\bma\times \sum_{s=\pm}\int \frac{d^2 \bp}{(2\pi\hbar)^2}f_s(\bx,\bp,t)\bs_s,\label{eq:tauST}
\end{eqnarray}
where $\bm{\tau}_{\rm{ST}}$ is the sum of all the separate spin torques we list below. We evaluate the integral in Eq. (\ref{eq:tauST}) up to first order in the damping parameter $\alpha$ and gradient of the magnetization and up to second order in the spin-orbit coupling strength $\lambda$. Note that we only include terms linear in the electric field, and that we take $\partial{\bma}/\partial t=0$. Taking into account this time-dependence gives rise to renormalization of damping and gyromagnetic ratio that we already discussed in the previous section.

In agreement with our phenomenlogical arguments [see Eqs. (\ref{eq:tst1},ref{eq:tst1p})], we obtain two spin torques that are zeroth order in the gradient of the magnetization which are given by
\begin{eqnarray}
\bm{\tau}^{(1)}&=&\frac{|e|m_e\lambda a^2}{\pi\hbar^2}\left(\frac{\Delta\tau_r}{2\hbar}-\alpha \frac{2\epsilon_F}{\Delta} \right) (\bE\times\uz)\times\bma;\quad\label{eq:tMZ}\\
\bm{\tau}^{(1\perp)}&=&\frac{|e|m_e\lambda a^2}{\pi\hbar^2} ((\bE\times\uz)\times\bma)\times\bma.\label{eq:tMZp}
\end{eqnarray}
These homogeneous SO induced spin torques where derived before.\cite{ManchonZhang2009,Kim2012} In case $\alpha=0$, we agree with Manchon and Zhang\cite{ManchonZhang2009,ManchonZhang2008} and with Kim \textit{et.al.}\cite{Kim2012} about the ratio between the reactive and dissipative torques. In addition, for $\alpha\neq0$, we find another contribution to these torques coming from spin relaxation. Note that the two torques given above form a perpendicular pair, one dissipative one reactive. In what follows we will group the torques into these pairs when both reactive and dissipative torques emerge to second order in SO coupling. 

All other torques are first order in the gradient of the magnetization. The first two torques we find are given by
\begin{eqnarray}
\bm{\tau}^{\rm{STT}}&=&-\frac{|e| \tau_r a^2}{ \pi \hbar^2}\left(\frac{\Delta}{2}+2\frac{ \alpha \hbar m_e \lambda^2}{\tau_r\Delta}-12m_e \lambda^2\frac{\epsilon_F}{\Delta}\right)\nonumber\\
&\quad&(\bE\cdot\nabla)\bma;\label{eq:stt}\\
\bm{\tau}^{\rm{STT}\perp}&=&\frac{|e|\alpha\tau_r a^2}{ \pi \hbar^2}\left(\epsilon_F  +\frac{m_e\lambda^2}{2}\left(7+\frac{4\epsilon_F^2}{\Delta^2}\left(3 +4m_z^2 \right) \right)\right)\nonumber\\
&\quad&\bma\times(\bE\cdot\nabla)\bma,\label{eq:sttp}
\end{eqnarray}
which are the well known STTs that also occur in systems with negligible SO coupling, see Eq. (\ref{eq:adiabatictorques}), and are due to the spin-polarized current in the direction of the electric field. The ratio of these two torques defines the $\beta$ parameter. We find that
\begin{equation}\label{eq:beta}
\beta = -\frac{2\alpha}{\Delta}\left(\epsilon_F + m\lambda_R^2\left(\frac{7}{2}+\frac{\epsilon_F^2}{\Delta^2}(30+8 m_z^2) \right)\right).
\end{equation}
In the previous section we showed that the current can be decomposed into three components. Several of the torques we find can be interpreted as the ordinary spin transfer torques [Eq. (\ref{eq:adiabatictorques})] with current response modified due to the SO coupling. First, we have the torques
\begin{eqnarray*}
-\frac{4|e|m_e \epsilon_F\lambda^2 a^2}{\pi\Delta^2\hbar} (\bma\cdot\uz)  ((\bE\times\uz)\cdot\nabla)\bma;\\
\frac{2|e|m_e\alpha \lambda^2 a^2}{\pi \Delta \hbar} (\bma\cdot\uz) \bma \times( (\bE\times\uz)\cdot\nabla)\bma,
\end{eqnarray*}
which are due to the anomalous Hall current, $j^i_{\rm{AH}}\equiv\sigma^{ij}_{\rm{AH}}E^j$, and are given to first order in gradients as
\begin{eqnarray}
\bm{\tau}^{\rm{AH}}&=&\mathcal{P}(\bj_{\rm{AH}}\cdot\nabla)\bma;\\
\bm{\tau}^{\rm{AH}\perp}&=&\mathcal{P}\frac{\alpha \Delta}{2\epsilon_F}\bma\times(\bj_{\rm{AH}}\cdot\nabla)\bma.
\end{eqnarray}
Two torques can be interpreted to be a generalization of the STTs coming from the anisotropic magnetoresistance response given by Eq.(\ref{eq:sigmaAMR}). These torques are
\begin{widetext}
\begin{eqnarray}
\bm{\tau}^{\rm{AMR}}&=&-\frac{2|e|m_e \lambda^2 a^2}{\pi\Delta\hbar}\left(\alpha-24\frac{\epsilon_F\tau_r}{\hbar} \right) ((\bE\times\uz)\cdot\bma) ((\bma\times\uz)\cdot\nabla)\bma;\\
\bm{\tau}^{\rm{AMR}\perp}&=&- \frac{|e|m_e\lambda^2\alpha\tau_r a^2}{\pi\hbar^2}(5+16\frac{\epsilon_F^2}{\Delta^2})((\bE\times\uz)\cdot\bma) \bma \times((\bma\times\uz)\cdot\nabla)\bma.
\end{eqnarray} 
\end{widetext}
The next torque, given by
\begin{equation}\label{eq:tauHall}
\bm{\tau}^{\rm{Hall}}=-\frac{|e|m_e\lambda^2\alpha\tau_r a^2}{2\pi\hbar^2}(1+4\frac{\epsilon_F^2}{\Delta^2})((\bE\times\bma)\cdot\nabla)\bma,
\end{equation}
has the symmetry of a STT due to a normal Hall resonse. This is not the normal Hall response because it is quadratic in the SO coupling parameter. In our description we did not include the normal Hall response of the system, due to the smallness of the effect.

The torques obtained up to this point could be interpreted as the known SO coupling induced spin torques for Eqs. (\ref{eq:tMZ}, \ref{eq:tMZp}) and the STTs [in Eqs. (\ref{eq:stt})--(\ref{eq:tauHall})] with a current response that is modified due to SO coupling. Now we will list the torques that cannot be interpreted as known current-induced torques. We have the pairs
\begin{widetext}
\begin{eqnarray}
\bm{\tau}^{\rm{a}}&=&\frac{2|e|m_e \lambda^2 a^2 }{\pi \Delta \hbar} \left(\frac{\epsilon_F\tau_r}{\hbar}-\alpha \right)(\bma\times(\bE\times\uz))^a \bma\times(\uz\times\nabla)m^a;\\
\bm{\tau}^{\rm{a}\perp}&=&-\frac{4|e|m_e\epsilon_F\lambda^2 a^2}{\pi\Delta^2\hbar}(\bma\times(\bE\times\uz))^a\bma\times( \bma\times(\uz\times\nabla))m^a,\, \rm{and}\\
\bm{\tau}^{\rm{b}}&=&-2\frac{\alpha |e|m_e \lambda^2 a^2 }{\pi \hbar}\frac{\tau_r}{\hbar}(\bE\times\uz)^a \bma\times(\uz\times\nabla)m^a;\\
\bm{\tau}^{\rm{b}\perp}&=&\frac{2|e|m_e\lambda^2 a^2}{\pi\Delta\hbar }\left(\alpha-\frac{\epsilon_F \tau_r}{\hbar} \right) (\bE\times\uz)^a \bma\times(\bma\times(\uz\times\nabla))m^a.
\end{eqnarray}
\end{widetext} 
We also have four torques that do not form reactive-dissipative pairs we list them below
\begin{eqnarray}
\bm{\tau}^{\rm{c}}&=&4\frac{|e|m_e\epsilon_F \lambda^2\tau_r a^2}{\pi\Delta\hbar^2}(\bma\cdot\nabla \bma\cdot\uz) \bma\times(\bE\times\uz);\\
\bm{\tau}^{\rm{d}}&=&-4\frac{|e|m_e\epsilon_F \lambda^2\tau_r a^2}{\pi\Delta\hbar^2}(\bE\times\uz)^a(\bma\cdot\nabla)m^a (\bma\times\uz);\quad \nonumber\\
\quad
\end{eqnarray}
\begin{eqnarray}
\bm{\tau}^{\rm{e}}&=&\frac{|e|m_e\lambda^2\alpha\tau_r a^2}{\pi\hbar^2}(1+4\frac{\epsilon_F^2}{\Delta^2})E^a (\bma\times\nabla) m^a;\\
\bm{\tau}^{\rm{f}}&=&\frac{3|e|m_e\lambda^2\alpha\tau_r a^2}{2\pi\hbar^2}(1+4\frac{\epsilon_F^2}{\Delta^2})\uz^a (\bma\times\uz)(\bE\cdot\nabla) m^a.\qquad
\end{eqnarray} 
Note that the above torques are of second order in $\uz$, and have therefore not been explicitly written down in Sec. \ref{sec:SymConSOC}.
The current-induced spin torques in this section are the central result of this paper. From the list of torques we presented here it is clear that the interplay of SO coupling and an inhomogeneous magnetization gives rise to many spin torques. In the next section we consider their effect on current-induced domain-wall motion.
\section{Domain-Wall Motion}\label{sec:DWM}
In this section we investigate the effect the spin torques have on current-induced domain-wall dynamics. We study the domain-wall dynamics by employing the one-dimensional rigid domain-wall model. Within this model the dynamics is captured by the collective coordinates of the wall which are its position $\rdw$ and central angle $\phidw$. We study three different realizations of domain walls summarized in Table \ref{tbl:walls}. Due to the SO coupling the current-driven motion of the three walls differs.
\begin{table}[b]
\caption{Magnetic anisotropy configuration and the corresponding domain wall structures.}
\begin{tabular}{l c c }\label{tbl:walls}
 & Easy Axis ($K$) & Hard Axis ($K_{\perp}$)\\
\hline
Bloch(z) & z&y\\
N\'eel(x) & x&z\\
Bloch(y)&y&x\\
\hline
\end{tabular}
\end{table}
In order to arrive at the equations of motion for the collective coordinates we describe the direction of the magnetization $\bma=\left(\cos\phidw\sin\thetadw, \sin\phidw\sin\thetadw, \cos\thetadw\right)$ using two angles $\thetadw$ and $\phidw$. We use $\thetadw(x,t)=2\arctan\left[\exp(x-\rdw)/\lambda_{\rm{dw}}\right]$ and $\phidw(t)$, where $\ldw=\sqrt{J/K}$ is the domain-wall width in terms of the exchange stiffness $J$ and the easy axis anisotropy $K$. The direction of the electric field is specified by the angle $\phi_E$ with the x-axis in the x-y plane. The known\cite{TataraKohno2004,*TataraKohno2006,ObataTatara2008} equations of motion for the collective coordinates $\rdw$ and $\phidw$ are augmented by terms obtained from the current-induced torques of the previous section. In the calculations we make use of the parameter values as given in Table \ref{tbl:parameters}. These parameters ar typical for metallic ferromagnets, and the value of the spin-orbit coupling is taken from Ref.[ \onlinecite{Miron2011,*MironNmat2011} ]. Furthermore, we give the results as a function of the critical field $ E_c$ and velocity $v_c$ for the case without SO coupling, which are defined as\cite{TataraKohno2004}
\begin{eqnarray}
v_c &=& \frac{K_{\perp}}{\hbar}\ldw;\\
E_c&=&\frac{v_c}{v_s^0},
\end{eqnarray}
where the spin velocity in absence of SO coupling is defined as
\[
v_s^0 = -\frac{|e|\tau_r\Delta a^2}{2\pi\hbar^2},
\]
which is the zero S) coupling ($\lambda\rightarrow0$) limit of Eq. (\ref{eq:stt}). In Eq. (\ref{eq:LLGr}) we introduced the renormalized Gilbert damping parameter $\alpha_G'$ which is the Gilbert damping paramater that will be measured in experiments. We expect that the Gilbert damping $\alpha_G$ for the magnetization and the damping $\alpha$ for the itinerant spins are of the same order of magnitude. 
\begin{table}
\caption{Parameters used in domain wall motion calculations.}
\label{tbl:parameters}
\begin{tabular}{l c l}
\hline
$\epsilon_F$&=&1 eV\\
$\Delta$&=&0.1  eV\\
$m\lambda^2$&=&9 meV\\
$\alpha_G'$&=&0.05\\
$\alpha$&=&0.05\\
$\tau$&=&30 fs\\
$\ldw$&=&10 nm\\
$a$&=& 0.3 nm\\
\hline
\end{tabular}
\end{table}
In the Appendix we give the equations of motion for the N\'eel(x) and Bloch(y) wall configurations. Here we explicitly address the Bloch(z) wall. 

The equations of motion for the collective coordinates are obtained  by inserting the Bloch(z) domain-wall \textit{ansatz}, as given above, into the equation of motion for the magnetization, see Eq. (\ref{eq:LLGr}). To get the equations of motion we take the inner-product with $\delta \bma_{\rm{Bloch(z)}}/\delta \rdw$, for one equation of motion and similar for  $\delta \bma_{\rm{Bloch(z)}}/\delta \phidw$. Subsequently we integrate those two equations over all space. The two equations of motion we obtain in this way are given below
\begin{widetext}
\begin{eqnarray}
\frac{\dot{r}_{\rm{dw}}}{\ldw} -\alpha_G'\dot{\varphi}_{\rm{dw}}-\frac{K_{\perp}}{\hbar}\sin 2\phidw&=&\left(\frac{\pi}{2}\tau^{(1)}+\frac{6\tau^{\rm{a}}+4\tau^{\rm{c}}}{6\ldw}\cos\phidw-\frac{\tau^{\rm{e}}}{3\ldw}\sin\phidw\right)E\cos(\phi_E-\phidw)\nonumber\\
&\quad&+\frac{2\tau^{\rm{b}\perp}-4\tau^{\rm{AMR}}}{6\ldw}E\sin(\phi_E-\phidw)\sin\phidw-\frac{\tau^{\rm{STT}}}{\ldw}E\cos\phi_E;
\end{eqnarray}
\begin{eqnarray}
\dot{\varphi}_{\rm{dw}}+\alpha_G'\frac{\dot{r}_{\rm{dw}}}{\ldw}&=&\left(\frac{\pi}{2}\tau^{(1\perp)}-\frac{\tau^{\rm{a}\perp}}{\ldw}\cos\phidw\right)E\cos(\phi_E-\phidw)\nonumber\\
&\quad& +\frac{4\tau^{\rm{AMR}\perp}+2\tau^{\rm{b}}}{3\ldw}E\sin(\phi_E-\phidw)\sin\phidw+\frac{2\tau^{\rm{f}}+3\tau^{\rm{STT}\perp}}{3\ldw}E\cos\phi_E,
\end{eqnarray}
\end{widetext}
The scalars $\tau^{(i)}$ are defined as the prefactors in front of the vector quantities of the torques in section \ref{sec:CItorques}. 
\begin{figure}[!t]
\centering
\includegraphics[width=1.0\linewidth]{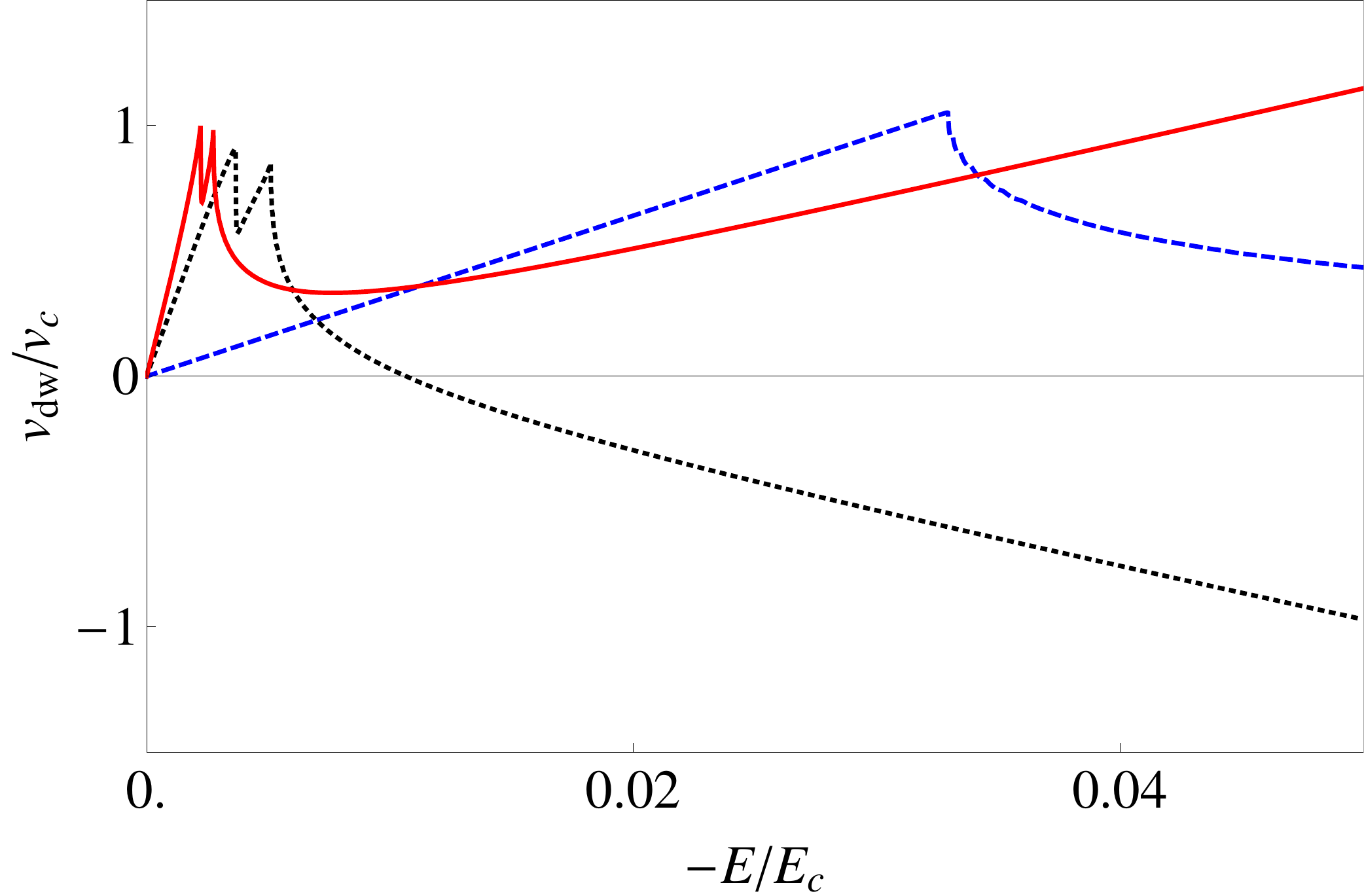}
  \caption{Average velocity of a Bloch(z) wall as a function of the applied field. The dashed (blue) line is the situation without spin-orbit coupling, the dotted (black) line shows the results with only the homogeneous SO torques, i.e. $\bm{\tau}^{(1)}$ and $\bm{\tau}^{(1\perp)}$ added. The solid line (red) shows the result of the solution of the equations of motion including all spin torques. The parameters used to obtain these results are given in Table \ref{tbl:parameters}}.
  \label{fig:Blochz}
\end{figure}
\begin{figure}[!t]
\centering
\includegraphics[width=1.0\linewidth]{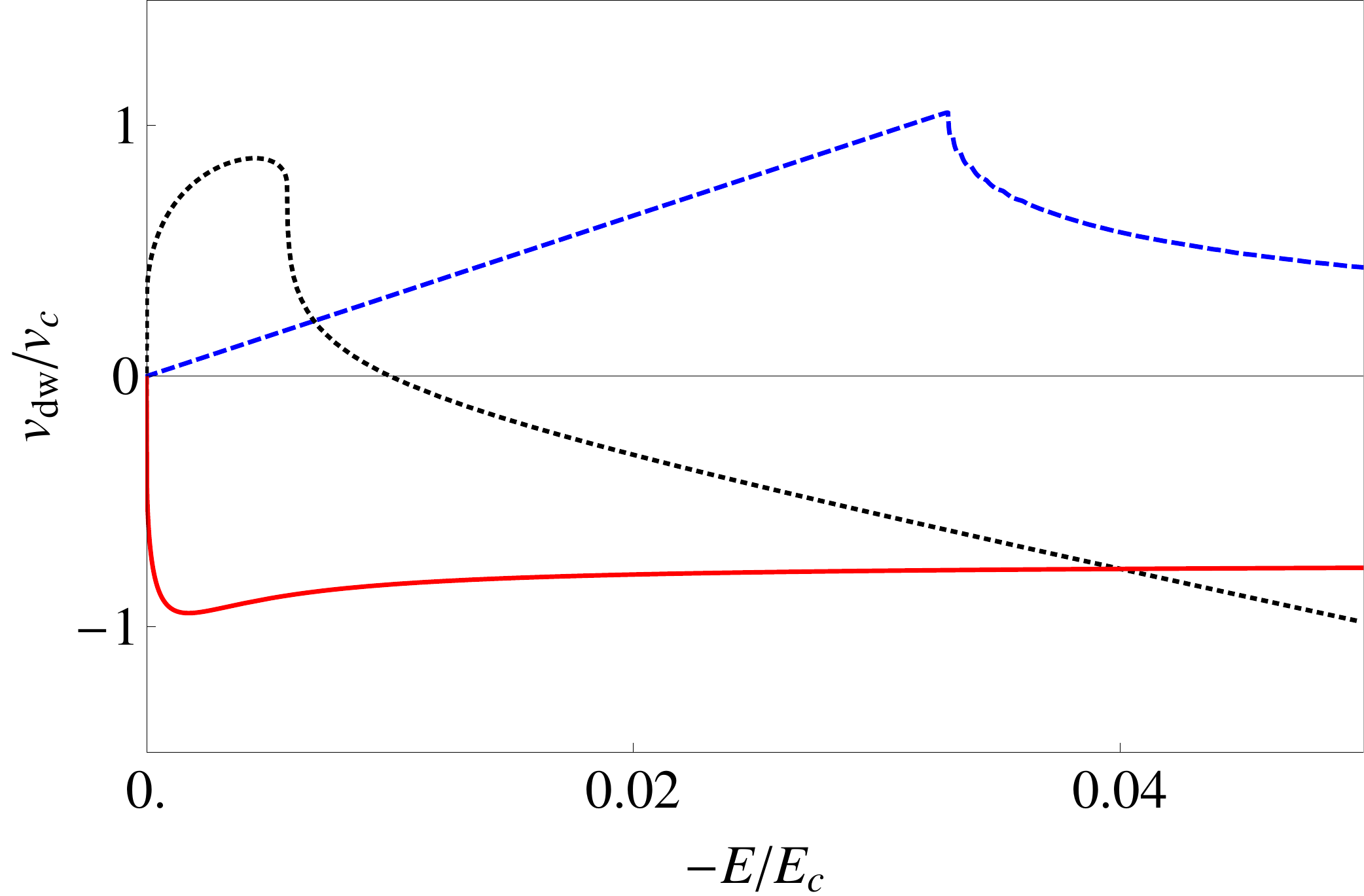}
  \caption{Average velocity of a N\'eel(x)-wall as a function of applied electrix field in the $x$-direction. Lines are as in Fig.\ref{fig:Blochz}. The equations of motion can be found in the Appendix.}
  \label{fig:Neelx}
\end{figure}
\begin{figure}[!t]
\centering
\includegraphics[width=1.0\linewidth]{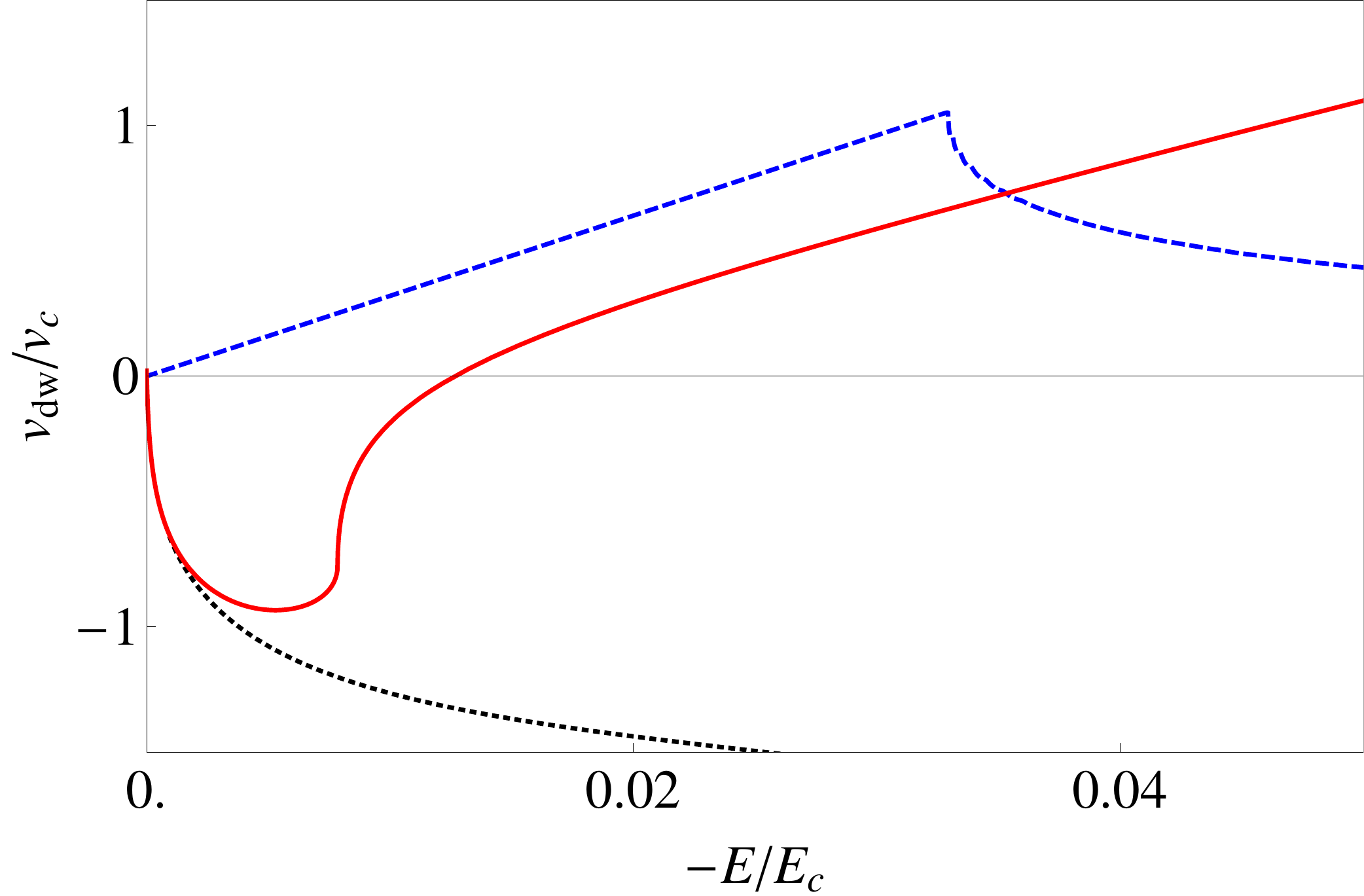}
  \caption{Average velocity of a Bloch(y)-wall as a function of applied electrix field in the $x$-direction. Lines are as in Fig.\ref{fig:Blochz}. The equations of motion can be found in the Appendix.}
  \label{fig:Blochy}
\end{figure}
In Fig. \ref{fig:Blochz} we show the average Bloch(z)-wall velocity as a function of the applied electric field in the x-direction. The boundary conditions for the current through the ferromagnet are such that only a current in the x-direction is present. In the figures we took $\phi_E=0$, since the off-diagonal contributions in the conductivity give rise to a small ($<1\%$ of the external field for the parameters used) voltage gradient in the y-direction. The average domain-wall velocity is defined as $v_{\rm{dw}}=\langle \dot{r}_{\rm{dw}} \rangle$, where the brackets denote a long-time average.

 From the results in Figs. \ref{fig:Blochz}--\ref{fig:Blochy} we see that the inclusion of spin torques due to the combined effect of do coupling and magnetization changes the domain-wall mobility $\mu_{\rm{dw}}=d v_{\rm{dw}}/dE$ completely as compared to the situation without these torques.

In Figs. \ref{fig:Neelx}, \ref{fig:Blochy} we show the results for the N\'eel(x) and Bloch(y) walls respectively. It is clear that also in this case the additional torques induce qualitatively different behaviour of the domain wall motion compared to the situation with only the torques induced by SO coupling for homogeneous magnetization.
\section{Discussion}
In this paper we considered current-induced torques in systems that have SO coupling and a textured magnetization. The effects of these torques on domain-wall motion have been investigated. We have shown that the effects of the interplay between the SO coupling and the gradients in the magnetization are qualitatively important for domain-wall dynamics. In particular, we showed that the inclusions of all torques typically changes the domain-wall mobility as compared to including only the spin transfer torques that occur at weak spin-orbit coupling and/or the homogeneous spin torques due to SO coupling. The results of this work may be used to discriminate between Rashba SO coupling and injection of a spin current via the spin Hall effect, because the latter will only show the homogeneous current-induced torques.

In this paper we considered Rashba SO coupling. Our results can be generalized straightforwardly to linear Dresselhaus SO coupling,\cite{Dresselhaus1955} which is linear in momentum too. For linear Dresselhaus coupling the dispersion of the carriers is the same as for Rashba coupling. The effective magnetization for the Dresselhaus SO coupling is given by $\bOm_{\rm{D}}(\bx,\bp)=\Delta\bma/2+\lambda_{\rm{D}}(-p_x,p_y,0)^T$. This means $\bp\times\uz\rightarrow(-p_x,p_y,0)^T$ when we go from the Rashba to the Dresselhaus coupling. The current-induced torques we found in Sec. \ref{sec:CItorques} involve factors $\bm{v}\times\uz$, where $\bm{v}$ is a vector. For clarity we consider $\bm{\tau}^{(1)}\propto (\bE\times\uz)\times\bma$ (given in Eq. (\ref{eq:tMZ})), for the Dresselhaus system the torque would be in the direction $(\bE\times\uz)\times\bma\rightarrow (-E_x,E_y,0)\times\bma$. In this way we obtain the results for the textured Dresselhaus ferromagnet. The results for combined Rashba-Dresselhaus SO coupling are less straightforward to obtain since the dispersion of the carriers changes.

Another obvious place to look for the appearance of addiational torques due to SO coupling would be in dilute magnetic semiconductor systems, where the effective Hamiltonian for the carriers also has strong SO coupling. In Ref. [\onlinecite{Culcer2009}] spin torques for the dilute limit are calculated for this system. In that work one of the current-induced torques is interpreted as an anisotropic dissipative STT. This anisotropic torque can as well be interpreted as the torque given by Eq. (\ref{eq:CIt7}). It would be very interesting to see which other torques would appear in those systems because the allowed spin torques would be listable since there is no symmetry breaking in the z-direction.

The reciprocal physical mechanism associated with current-induced torques are currents driven by non-equilibrium magnetization dynamics, often referred to as spin-motive forces. We obtain these using the Onsager reciprocal relations.\cite{WongTserkovnyak2009} We do this via the linear response matrix
\begin{equation*}
\begin{pmatrix} \dot{m}^i\\ j_c^i \end{pmatrix} =\begin{pmatrix} m^k\epsilon^{ijk} & L_{\rm{cit}}^{ij}(\bma,\uz,\nabla\bma) \\L_{\rm{smf}}^{ij}(\bma,\uz\nabla\bma) &\sigma^{ij}(\bma,\uz,\nabla\bma) \end{pmatrix}\cdot\begin{pmatrix} H_{\rm{eff}}^j\\ E^j \end{pmatrix},
\end{equation*}
where $L_{\rm{cit}}^{ij}(\bma,\uz,\nabla\bma)$ is the ($3\times3$) matrix that gives the current induced torques as defined in Eq. (\ref{eq:Lcit}) and $L_{\rm{smf}}^{ij}(\bma,\uz,\nabla\bma) $ gives the spin motive forces. These two matrices are related via Onsager reciprocity which yields
\begin{equation*}
L_{\rm{cit}}^{ij}(\bma,\uz,\nabla\bma) = L_{\rm{smf}}^{ji}(-\bma,\uz,-\nabla\bma).
\end{equation*}

In future work we intend to explore these spin motive forces in more detail. Another interesting direction for future research is the inclusion of thermal gradients and heat currents.
\acknowledgements
It is a pleasure to thank Arne Brataas and Dima Pesin for useful remarks.
This work was supported by the Stichting voor Fundamenteel
Onderzoek der Materie (FOM), the Netherlands
Organization for Scientifc Research (NWO), and by the
European Research Council (ERC).
\bibliography{SpinOrbitSym.bbl}
\appendix
\section{Different domain-wall configurations}
In this appendix we give the equations of motion for the N\'eel(x) and Bloch(y) domain-wall configurations. The magnetic anisotropy configuration corresponding to these different walls is given in Table \ref{tbl:walls}.
\subsection{N\'eel(x) Wall}
The N\'eel(x) wall is parameterized as $\bma = (\cos(\theta(\bx,t)),\cos\phi(\bx,t)\sin(\theta(\bx,t)),\sin\phi(\bx,t)\sin(\theta(\bx,t)) )^T$. The equations of motion are obtained as explained in Sec. \ref{sec:DWM} of the main text. The equations of motion for the collective coordinates are given by
\begin{widetext}
\begin{eqnarray*}
\frac{\dot{r}_{\rm{dw}}}{\ldw} -\alpha_G'\dot{\varphi}_{\rm{dw}}&=&\frac{K_{\perp}}{\hbar}\sin 2\phidw-\left(\tau^{(1\perp)}+\frac{1}{3\ldw}\tau^{\rm{e}} \cos^2\phidw+\frac{\pi}{4\ldw}\left(\tau^{\rm{AH}}-\tau^{\rm{b}}+\tau^{\rm{Hall}}\right)\sin\phidw \right)E\sin \phi_E \\
&-&\left(\frac{\pi}{2}\tau^{(1)}+\frac{1}{3\ldw}\left(-3\tau^{\rm{AMR}}+\tau^{\rm{b}\perp}+\tau^{\rm{d}}\right)\cos^2\phidw-\frac{1}{3\ldw}(3\tau^{\rm{a}}+\tau^{\rm{c}})\sin^2\phidw\right)E\cos\phi_E,\\
\end{eqnarray*}
\begin{eqnarray*}
\dot{\varphi}_{\rm{dw}}+\alpha_G'\frac{\dot{r}_{\rm{dw}}}{\ldw}&=&\frac{1}{\ldw}\Bigg( \frac{\pi}{2}\left( \frac{1}{2}\tau^{\rm{e}}-\ldw\tau^{(1\perp)} \right)\sin\phidw\\
&\qquad&-\frac{1}{3}\left(\left( 2\tau^{\rm{AMR}}+\tau^{\rm{b}} \right)\cos^2\phidw-(-3\tau^{\rm{a}\perp}+\tau^{\rm{f}})\sin^2\phidw\right) +\tau^{\rm{STT}\perp} \Bigg)E\cos\phi_E\\
&+&\left(\tau^{(1)}+\frac{\pi}{48\ldw}\left( 4 \tau^{\rm{AH}\perp}-4\tau^{\rm{b}\perp}-\tau^{\rm{c}}-\tau^{\rm{d}} \right)\sin \phidw  \right)E\sin\phi_E.
\end{eqnarray*}
\subsection{Bloch(y) Wall}
For the Bloch wall the magnetization is parameterized as $\bma = (\cos\phi(\bx,t)\sin(\theta(\bx,t)),\cos(\theta(\bx,t)),\sin\phi(\bx,t)\sin(\theta(\bx,t)) )^T$. The equations of motion are
\begin{eqnarray*}
\frac{\dot{r}_{\rm{dw}}}{\ldw} -\alpha_G'\dot{\varphi}_{\rm{dw}}-\frac{K_{\perp}}{\hbar}\sin 2\phidw&=&-\frac{1}{3\ldw}\left(\tau^{\rm{AMR}}+2\tau^{\rm{b}}-3\tau^{\rm{STT}\perp}-\tau^{\rm{f}}\cos^2\phidw+3\ldw\tau^{(1)}\right)E\cos\phi_E\\
&-&\left(\frac{2}{3\ldw}+\frac{\pi}{4\ldw}\left( \tau^{\rm{AH}}-\tau^{\rm{a}\perp}+\tau^{\rm{Hall}}+2\ldw\tau^{(1)} \right)\cos\phidw  \right)E\sin\phi_E,\\
\end{eqnarray*}
\begin{eqnarray*}
\dot{\varphi}_{\rm{dw}}+\alpha_G'\frac{\dot{r}_{\rm{dw}}}{\ldw}&=&-\frac{1}{3\ldw}\left( \tau^{\rm{AMR}\perp}+2\tau^{\rm{b}}+3\ldw\tau^{(1)}-3\tau^{\rm{STT}\perp}-\tau^{\rm{f}}\cos^2\phidw \right)E\cos\phi_E\\
&\quad&+\frac{\pi}{32\ldw}\left(8\tau^{\rm{a}}+8\tau^{\rm{AH}\perp}+\tau^{\rm{c}}+\tau^{\rm{d}}-16\ldw\tau^{(1\perp)} -(\tau^{\rm{c}}+\tau^{\rm{d}})\cos2\phidw \right)\cos\phidw E\sin\phi_E.
\end{eqnarray*}
\end{widetext}
\end{document}